\documentclass[doublecol]{epl2} 

\def\be{\begin{equation}}
\def\en{\end{equation}}    

\newcommand{\bi}[1]{\mbox{\boldmath$#1$}}

\def\p{\partial}
\def\bea{\begin{eqnarray}}
\def\ena{\end{eqnarray}}

\def\asigma{\stackrel{\leftrightarrow}{\sigma}}
\def\aPi{\stackrel{\leftrightarrow}{\Pi}}

\def\gs{> \kern -12pt \lower 5pt \hbox{$\displaystyle{\sim}$}}
\def\ls{< \kern -12pt \lower 5pt \hbox{$\displaystyle{\sim}$}}

\title{Droplet evaporation 
in one-component fluids: Dynamic van der Waals theory }
\shorttitle{Droplet evaporation} 

\author{Ryohei  Teshigawara  and Akira Onuki}
\shortauthor{Teshigawara and  Onuki}

\institute{                    
Department of Physics, Kyoto University, 
Kyoto 606-8502, Japan
}

\abstract{
In a one-component fluid, we investigate  evaporation of a 
small axysymmetric liquid  droplet in the partial 
wetting condition on a heated wall  
 at $T\sim 0.9 T_c$.  In  the dynamic van der Waals theory 
(Phys. Rev. E  {\bf 75},  036304 (2007)), we take into account 
the latent heat transport from liquid to gas  upon evaporation.  
Along  the gas-liquid interface, the temperature 
is nearly equal to the equilibrium coexisting 
temperature away from the substrate, 
but it rises sharply to the wall  
temperature close to the substrate.   
On an isothermal substrate,  
evaporation  takes place  mostly on a narrow 
interface region near the contact line in a late stage, 
which is a  characteristic feature in 
one-component fluids. }

\begin{document}

\maketitle


\section{Introduction}
The  wetting dynamics 
has been mostly studied for imvolatile liquids 
and is  not  well understood  for  volatile liquids \cite{PG}. 
A well-known example is  evaporation of a liquid droplet 
on a heated  substrate. 
Here first-order phase transition 
from liquid to gas occurs  on the interface, 
where latent heat is carried  away from 
the interface with gas flow.  
For a liquid droplet in air 
the  radius $r_c(t)$ has been observed 
  to decrease as $r_c(t) \sim (t_0-t)^a$ with $a \cong 0.5$ 
 until it vanishes at time $t_0$  
\cite{Nature,Caza,Bonn,Butt}. 
Detailed information of 
evaporation   and contact-line motion 
has been provided by various theoretical approaches 
\cite{Davis,Koplik,Yeomans,Savino,Larson,Nikolayev,NikolayevNEW}.
In near-critical one-component fluids, 
 in particular, a  bubble in liquid 
was observed to be attracted  to a heated wall 
even when the wall was  wetted 
by liquid in equilibrium 
\cite{Hegseth}. In one-component fluids,   
the contact angle  decreases 
from zero to an apparent finite value 
in heat flux.

In this Letter, we numerically 
study  evaporation  of a droplet  
in one-component fluids  in the axisymmetric geometry. 
 As an efficient  numerical method,  we 
use  the dynamic van der Waals model 
 \cite{OnukiV}, which is a phase field  model of fluids 
with inhomogeneous temperature. 
The pressure $p$ outside a droplet (or bubble) 
is nearly homogeneous  so that  
the interface temperature should   be 
 close to the equilibrium coexisting 
(saturation) temperature $T_{\rm cx}(p)$ even 
in heat flux  \cite{OnukiV,Kanatani}. 
Thus,  near the contact line on a 
heated or cooled wall, 
a steep temperature variation 
and  a large heat flux 
should appear, as theoretically 
 studied by Nikolayev {\it et al.} 
 \cite{Nikolayev,NikolayevNEW} and as measured  by 
H$\ddot{\rm o}$hmann and Stephan \cite{Stephan}. 
Therefore, the hydrodynamics 
is singular around the contact line 
in heat flux.  This aspect 
has not yet been well studied in the literature. 
On the other hand, in multi-component fluids, 
the interface temperature changes 
on the scale of the droplet size and 
evaporation should take place 
all over the surface.

\section{Dynamic van der Waals model} 

We examine  the gas-liquid phase transition 
in nonstationary, inhomogeneous temperature $T({\bi r},t)$. 
We  start with 
the entropy functional ${\cal S}_b$ 
dependent on the number density $n({\bi r},t)$ 
and the internal  energy density $e({\bi r},t)$ 
and introduce $T({\bi r},t)$ by the functional derivative 
$1/T=(\delta {\cal S}_b/\delta n)_e$. 
We assume that ${\cal S}_b=\int d{\bi r}\hat{S}$ is 
the space integral of
the entropy density 
$\hat{S}$   consisting   of  regular 
 and  gradient parts  as 
\be 
\hat{S}= ns(n,e)-
\frac{1}{2}C | \nabla n |^2, 
\en 
where  $s$  is 
the entropy per particle  dependent    
on  $n$ and  $e$. 
The second term is the  gradient entropy 
density, 
which is negative and 
is imortant near the interface. 
The coefficient  $C$ is assumed to be a constant.  
In the van  der Waals theory, 
 fluids are characterized by 
the molecular volume $v_0$ and  the  
attractive pair interaction energy $\epsilon$ and
 the entropy $s$ is written as 
\be 
 s =  k_{B}      \ln [(e/n+ \epsilon v_0n)^{3/2}
  ({1}/{v_0n}-1 )]  +s_0, 
\en 
where $s_0/k_B= \ln [v_0 ({m/3\pi\hbar^2})^{3/2}]+5/2$ 
with $m$ being the molecular mass. 
In this Letter, we assume that 
the fluid internal energy ${\cal E}_b$ 
is the space integral of $e$. Then 
we have the usual relation 
$ 
1/T=
n(\p s/\p e)_n$, which yields the well-known expression  
 $e=3nk_BT/2-\epsilon v_0n^2$. 
More generally, 
we may assume 
the form ${\cal E}_b=\int d{\bi r}
[e+K | \nabla n |^2/2]$, where 
the second term represents 
the gradient energy 
density\cite{OnukiV}. 
In this Letter, we set $K=0$ for simplicity.

We set up the  hydrodynamic equations 
from the principle of 
positive entropy production in nonequilibrium. 
No gravity is assumed. 
The mass density $\rho=mn$ obeys the continuity equation,  
\be 
\frac{\p}{\p t} \rho = - \nabla \cdot(\rho{\bi v}).
\en  
The momentum density $\rho {\bi v}$ 
and the total energy density 
$e_{\rm T}\equiv  e+\rho {\bi v}^2/2$  
are  governed by \cite{Landau} 
\bea 
\frac{\p}{\p t}\rho {\bi v}
&=&-\nabla \cdot (\rho  {\bi v}{\bi v}+
 {\aPi}-{\asigma} ), 
\\
\frac{\p }{\p t} e_{\rm T} &=&- \nabla\cdot
\bigg[e_{\rm T}{\bi v} 
+({\aPi}- {\asigma})\cdot{\bi v}-\lambda\nabla T\bigg]. 
\ena  
Here ${\aPi}=\{\Pi_{ij}\}$ is the reversible stress 
tensor  consisting of   the  van der Waals pressure $p
=k_BTn/(1-v_0n)-\epsilon v_0 n^2$  
and the gradient contribution as 
\be
\Pi_{ij}= {p}\delta_{ij}   
-CT\bigg[ (n\nabla^2 n+ \frac{1}{2}|\nabla n|^2)\delta_{ij} 
-\nabla_i n \nabla_j n\bigg]. 
\en 
The ${\asigma}= \{{\sigma}_{ij}\}
=\eta(\nabla_i v_j+\nabla_j v_i)  +
(\zeta-2\eta/3) \nabla \cdot {\bi v}\delta_{ij}$ 
is the viscous stress tensor in terms of the 
shear viscosity $\eta$ and the bulk viscosity 
$\zeta$. Hereafter $\nabla_i=\p/\p x_i$ 
with $x_i$ representing $x$, $y$, and $z$. 
The $\lambda$ in eq.~(5) is the thermal conductivity.

We note that  $\Pi_{ij}$ satisfies  
$\sum_j \nabla_j(\Pi_{ij}/T)= 
n\nabla_i\hat{\mu} 
-e\nabla_i T^{-1}$, where 
$\hat{\mu}$ is the generalized 
chemical potential including the gradient contribution. 
If $C$ is assumed to be a constant, $\hat{\mu}$ reads   
\be 
\hat{\mu}= T(\delta{\cal S}_b/\delta n)_e= 
\mu -TC \nabla^2n, 
\en 
where  ${\cal S}_b=\int d{\bi r}\hat{S}$ 
is the entropy functional and $\mu=\mu(T,n)$ is 
 the usual chemical potential.  
 In equilibrium 
the stress balance  $\sum_j \nabla_j\Pi_{ij}=0$ 
is equivalent to the homogeneity of $\hat\mu$. 
The interface profile $n=n(x)$ is obtained  from 
$\mu(T,n)-CT d^2  n/dx^2=
\mu_{\rm cx}(T)$,  where 
$\mu_{\rm cx}(T)$ is the chemical  potential 
in two phase  coexistence. 
This interface equation was derived by 
van der Waals\cite{vander}.

In our simulation, however, 
we solved the equation for the entropy density 
$\hat{S}$ in eq.~(1), 
\be
\frac{\p\hat{S}}{\p t}+\nabla\cdot \bigg[
\hat{S}{\bi v}-Cn(\nabla\cdot {\bi v})\nabla n
-\frac{\lambda}{T}\nabla T
\bigg]= 
\frac{\dot{\epsilon}_v+\dot{\epsilon}_\theta}{T}, 
\en
together with  the continuity and  
momentum equations (3) and (4).  
The right hand side of eq.~(8) 
is the  nonnegative-definite 
entropy  production rate with 
\be   
\dot{\epsilon}_v= \sum_{ij}\sigma_{ij}\nabla_j v_i,
\quad 
\dot{\epsilon_\theta}= \lambda(\nabla T)^2/T.
\en  
In the literature on 
simulations of two phase fluids\cite{para,Jamet1,Daru}, 
it is well-known that
a small  velocity field remains  nonvanishing 
around the interface of a droplet  after 
long times even without heat input 
from the boundaries.  
It is  an artificial {\it parasitic} flow 
and its magnitude depends 
on the discretization method. In our previous 
work  \cite{OnukiV}, we integrated 
the energy equation (5) and 
fig.~3 there is    affected by such a parasitic flow. 
Instead, if we use  the entropy equation (8), 
the entropy production rate 
tends to zero or  $\nabla_iv_j\rightarrow 0 $ 
and $\nabla_iT\rightarrow 0$
 at long times 
if there is no heat flow from outside. 
Thus, with our new method, 
 equilibrium can be reached 
around the interface region.

In passing,   $s_0$ in eq.~(2) disappears in eq.~(8)  
owing to the continuity equation (4), so  
It is  an arbitrary constant. 

\section{Numerical method}

\begin{figure}
\begin{center}
\includegraphics[width=0.75\linewidth]{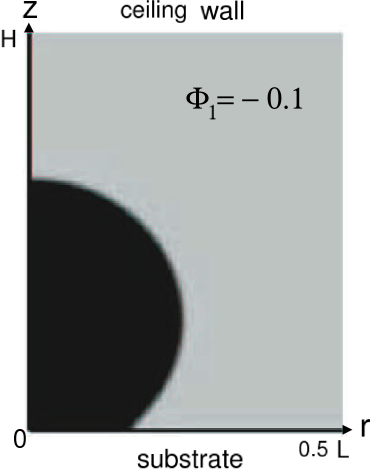}
\end{center}
\caption{\protect
Axisymmetric liquid droplet (in black) on the 
substrate in gas (in gray) 
for  $\Phi_1=-0.1$ and  $T=0.875T_c$
in equilibrium in a cylindrical cell 
($0<z<H$ and $0<r<L$). 
Here the region $r<L/2$ 
is shown. A thermallly insulating side wall 
is at $r=L$.  In this work $H=400\Delta x=200\ell$ and 
$L=700\Delta x=350\ell$.}
\end{figure}

We assume a cylindrical cell 
with $0<z<H$ and $0 < r=\sqrt{x^2+y^2}<L$ filled with 
a fluid. The system volume 
$\pi L^2H$ is thus fixed. In this  axisymmetric geometry, 
we assume that all the variables depend only on $z, r,$ and $t$. 
We integrated  the dynamic equations on a two-dimensional 
lattice with  $H=400\Delta x$ and $L=700\Delta x$, 
where  $\Delta x=\ell/2$ is the mesh 
size of the integration 
with  
\be 
\ell=(C/2k_B v_0)^{1/2}
\en 
in terms of $C$ in eq.~(1) and 
$v_0$ in the van der Waals theory. 
For  $C \sim k_Bv_0^{5/3}$ we have $\ell\sim v_0^{1/3}$, but 
$C$ remains arbitrary in our theory.  
The interface width in the following simulations 
is then  of order $6\Delta x$ with 
$\Delta x=\ell/2$ and  we obtained 
almost the same results 
even for $\Delta x=\ell/4$. 
We assumed the linear density-dependence 
of the transport  coefficients  as 
$\eta=\zeta= \nu_0 mn $  and $\lambda= k_B\nu_0 n$, though 
they are  crude approximations  \cite{Bird}. 
The  kinematic viscosity $\nu_0=  \eta/\rho$ is 
assumed to be a constant.   Space and time will be 
measured in units of $\ell$ and 
\be
\tau_0= \ell^2/\nu_0=C/2k_B v_0\nu_0, 
\en 
respectively.   
Away  from the criticality, 
the thermal diffusivity $D_T= \lambda/ C_p$  
is of order $\nu_0$ (where $C_p$ 
is the  isobaric specific heat per unit volume) 
or the   Prandtl number 
$Pr= \nu_0/D_T$ is of order unity. 
Hence  $\tau_0$ is the 
thermal relaxation time on the  scale of $\ell$.   
There   arises a dimensionless number     
$\sigma\equiv \nu_0^2m/\epsilon\ell^2$   
and we set $\sigma=0.06$  in this work. 
Then $\nu_0= (0.06\epsilon/m)^{1/2}\ell$ or 
$C=k_Bv_0\nu_0^2m/0.03\epsilon$. 
The velocity field $\bi v$ 
vanishes on all the boundaries. 
We control the boundary temperatures  
 at $z=0$  and $H$, written as 
$T_0$ and $T_H$,  while the side wall at $r=L$ 
is thermally insulating 
or $(\p T/\p r)_{r=L}=0$. 
On the substrate $z=0$ we 
 imposed the boundary condition \cite{Yeomans},  
\be 
\frac{\p n}{\p z}= -  \Phi_1/v_0\ell,
\en 
where  $\Phi_1$ arises from  the short-range 
 interaction between the fluid and the wall 
assumed to be smooth \cite{PG}.  We  set 
$ {\p n}/{\p z}=0$ at  $z=H$ 
and $ {\p n}/{\p r}= -0.1/v_0\ell$ at  $r=L$.

We first placed a hemisphere liquid droplet 
with radius 100 at  $T=0.875T_c$, where 
the liquid and gas densities 
were those on the coexistence curve 
given by $n_\ell=0.580/v_0$ and 
$n_g= 0.122/v_0$, respectively.  
We then waited a time interval of $4200$ 
to let the system relax to the 
true equilibrium, where  $T=0.875T_c$ 
throughout the system and   
${\bi v}={\bi 0}$ even around  the interface, 
and the pressure difference between the two phases 
satisfied   the Laplace law. 
The liquid and gas densities attained are 
$0.581/v_0$ and $0.123/v_0$, respectively, which are 
slightly  different from the initial 
values due to  the surface tension effect. 
In fig.~1, we show such a state with 
 $\Phi_1=-0.1$, where the contact angle is 
 $135^\circ$.   At   $T=0.875T_c$, 
the contact angle vanishes  for  $\Phi_1\cong 0.12$ 
and the wall is completely wetted by liquid for 
larger $\Phi_1$. We determine the interface 
position by $n(r,z)=(n_\ell+n_g)/2$. 
As $z\rightarrow 0$ we obtain the position of the 
contact line  $r=r_c$.

\section{Heating  a liquid droplet}
 
\begin{figure}
\begin{center}
\includegraphics[width=0.75\linewidth]{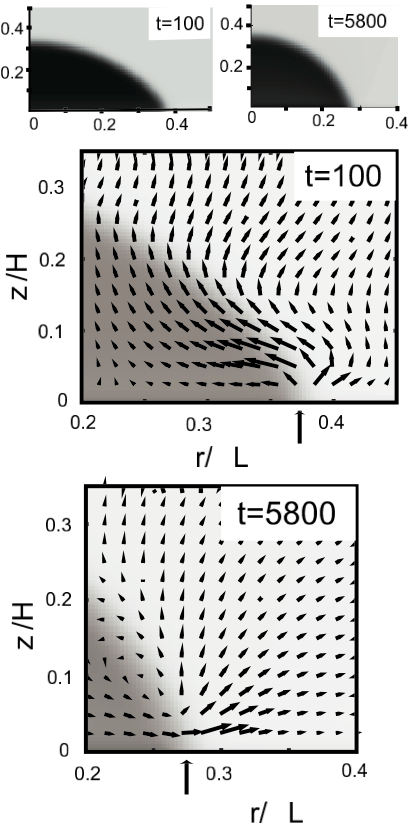}
\end{center}
\caption{\protect
Droplet shape  (top)  
and velocity field (middle 
and bottom) for $\Phi_1=0.05$ 
at $t=100$ and $5800$ on a heated substrate with  
the darker regions representing liquid. 
Evaporation strongly  takes place around the contact 
line, as can be seen from the velocity field (arrows). }
\label{fig:velocity_heat}
\end{figure}

\begin{figure}
\begin{center}
\includegraphics[width=0.75\linewidth]{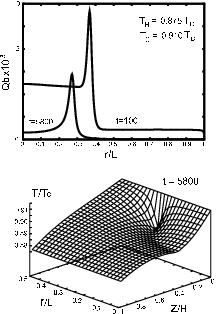}
\end{center}
\caption{\protect
Upper plate: 
Heat flux $Q_{\rm b}(r,t)$ on a heated 
 substrate in units of $\epsilon\ell/v_0\tau_0$ 
at $t=100$ and $5800$ with  a   sharp maximum 
at the contact line $r=r_{\rm c}(t)$. 
Lower plate: Profile of $T(r,z,t)/T_c$ at $t=5800$.  
It sharply dropls to  $0.8815$ near the contact line 
($r= r_c(t)$ and $z= 0$), where 
evaporation occurs strongly.
}
\end{figure}

We  prepared   an equilibrium 
state with $\Phi_1=0.05$  in the partial wetting condition, where 
the  contact angle is   $57^\circ$. 
We  then raised the substrate 
temperature $T_0$  to $0.91T_c$, while   
the temperature  $T_H$ at the ceiling was 
 unchanged from  $0.875T_c$. 
We set $t=0$ at this temperature change. 
The $\Phi_1$ was kept fixed. 
The  contact angle  increased up to $67^\circ$  around $t=1800$ 
in accord with the experiment \cite{Hegseth}, 
but it  slowly decreased  afterwards 
being equal to  $60^\circ$ at $t=10800$. 
(These two times correspond to the beginning and 
middle of an  late stage of evaporation  
in fig.~4  below.) 
The droplet assumed a cap-like shape 
until  it disappeared.

In fig.~2, 
the droplet shape and the velocity field are shown at 
$t=100$ in an early stage and at $t=5800$ 
in a late stage.  Evaporation is   
taking place strongly in the vicinity 
of  the contact line $r=r_c(t)$. 
In the upper panel  of fig.~3, we show 
the heat flux on  the substrate, 
\be 
Q_{\rm b}(r,t)= -\bigg(\lambda \frac{\p T}{\p z}\bigg)_{z=0}
\en 
at these two  times.  It   exhibits a sharp  
peak at  $r=r_c(t)$,  where the  wall 
supplies excess  heat needed for evaporation. 
This behavior is in accord with the 
theoretical result by Nikolayev {\it et al.}\cite{Nikolayev}
In the lower panel of fig.~3 at $t=5800$, 
the temperature $T$ is nearly 
constant along the interface, as 
in our previous simulation \cite{OnukiV}. 
It is nearly equal to the coexistence 
temperature $T_{\rm cx}(p)$,  
where $p$ is the pressure   homogeneous 
outside the interface region, 
as  anlytic calculations 
demonstrated \cite{Kanatani}.

In our simulation,  $T$ 
sharply drops from $T_0$ to $T_{\rm cx}(p)$ near 
the contact line  on the scale of 
 the interface thickness 
$\xi$,  while $T$ gradually 
decreases outside the contact line  region $|r-r_c|\gg \xi$. 
 We  confirmed this results 
even  for other  $T_0$ in the 
partial wetting condition. 
The heat flux near  the 
contact line is then given by 
$Q_{\rm con} =\lambda_\ell \Delta T/\xi$, 
where  $\Delta T= 
T_0-T_{\rm cx}$ and  $\lambda_\ell$ 
is the thermal conductivity of liquid. 
This heat flux may be  equated with 
the convective latent heat flux 
$n_g T_{\rm cx}\Delta s v_g$ 
in the gas region, where $n_g$ is the 
gas density, $\Delta s$ is the entropy difference per 
particle, and $v_g$ is the gas velocity 
near  the contact lne.  
The convection dominates over the thermal 
diffusion in this region.  Therefore, 
\bea 
v_g &\sim&
 \lambda_\ell \Delta T/
(\xi n_g T_{\rm cx}\Delta s) \nonumber\\
&\sim&  \nu_0 n_\ell \Delta T/
\xi n_g T_{\rm cx}. 
\ena 
In the second line we 
set  $\lambda_\ell=k_B\nu_0 n_\ell$ and 
$\Delta s \sim k_B$. 
This  estimation is consistent 
 with the numerical  
 values of $v_g$ in fig.~2. 
We also changed $\Delta T$ 
and confirmed the linear relationship  
$v_g\propto \Delta T$ in the partial wetting condition.

In  fig.~4,  we display   the 
radius of the contact 
line $r_{\rm c}(t)$ versus $t$.  
In the early stage $t\ls 10^3$,  
  $r_c(t)$ 
decreases rapidly, where 
evaporation takes place strongly 
all over the surface  (see fig.~5 below). 
In the late stage  $t\gs 10^3$,  it 
decreases  algebraically as  
\be 
r_{\rm c}(t) \propto (t_0-t)^{0.42}, 
\en  
until it disappears at $t_0 \cong  
2\times 10^4$. The 
exponent $0.42$ in eq.~(15)  is smaller than 
the exponent   0.5 
for macroscopic droplets in air \cite{Nature,Caza,Bonn}. 
The droplet  volume 
decreased as  $(t_0-t)^{1.12}$ and 
the average droplet density $\bar n(t)$ 
slowly decreased   
roughly as $\bar{n}-n_g \propto (t_0-t)^{0.08}$.  
The latter is because
 the droplet interior is  gradually heated,  
as can be seen from 
the inhomogeneous temperature profile in fig.~3.  
The interior density 
approaches  the gas density  before its disappearance. 
The droplet interior thus changes in a complicated manner 
and we cannot present clear explanations of  the exponents 
0.42, 1.12, and 0.08 at present.

\begin{figure}
\begin{center}
\includegraphics[width=0.75\linewidth]{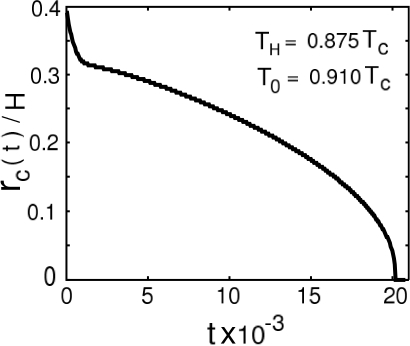}
\end{center}
\caption{\protect
Time evolution of the contact line radius 
$r_{\rm c}(t)$ of an evaporating droplet. 
For $t\gs 10^3$ it 
 may be fitted to eq.~(15). For $t\ls 10^3$ 
it decreases  rapidly due to 
enhanced evaporation.}
\end{figure}

\begin{figure}
\begin{center}
\includegraphics[width=0.75\linewidth]{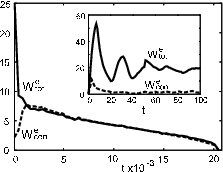}
\end{center}
\caption{\protect
Evaporation rate on all the interface $W^e_{\rm tot}$ 
in eq.~(18) and  that  $W^e_{\rm con}$ in the vicinity 
of the contact line   $r_c-6\ell<r<r_c$ in eq.~(19)   
in units of $\ell^3/v_0\tau_0$  vs $t$  
 until disappearance of the droplet. 
Evaporation occurs all over the interface 
in an early stage after heating the substrate, 
but it mostly occurs 
at the contact line 
in a late stage. 
Inset: evaporation rates 
 in the early stage, where 
sound propagation in liquid 
gives rise to oscillatory relaxation.
}
\end{figure}

\section{Evaporation rate} 

We next examine the  evaporation rate. 
To this end   we introduce the mass flux 
through the interface, 
\be
J =n ({\bi v}-{\bi v}_{int})\cdot{\bi \nu}, 
\en 
where ${\bi \nu}=- |\nabla n|^{-1}\nabla n$ 
 is the normal unit vector at the surface from liquid to gas 
and ${\bi v}_{int}\cdot{\bi \nu} 
(\cong  dr_c(t)/dt)$ is the interface velocity. 
If $J$ is regarded as a function 
of the coordinate along the normal direction $\bi \nu$, 
it is continuous through 
the interface from the mass conservation, 
while $n$ and ${\bi v}\cdot{\bi \nu}$ 
change discontinuously. Thus we may determine 
$J=J(r,t)$ on the interface 
 as a function of $r$ at each time. 
In the thin interface limit  $J$ is related to the 
discontinuity of the heat flux 
$-\lambda\cdot\nabla T$ 
as 
\be 
T\Delta s J- {\bi \nu}\cdot[ 
(\lambda\nabla T)_{\rm gas}-(\lambda\nabla T)_{\rm liq}]=0,
\en    
from the energy conservation at the interface, where 
the subscript gas (liq) denotes the value 
in the gas (liquid) side close to the interface.   
The  total evaporation rate $W_{\rm tot}^e$ 
is the surface integral  of $J$. 
The surface area in the range $[r,r+dr]$ 
is $2\pi dr r/\sin\theta$, where  $\theta$ is  the angle 
between  $\bi \nu$ and the $r$ axis.   
Thus, 
\be
W_{\rm tot}^e =2\pi \int_{0}^{r_{\rm c}} dr {r}J/{\sin\theta}    .  
\en 
The particle number within the droplet $N_d$ 
decreases in time as 
$dN_d/dt= -W_{\rm tot}^e$. The droplet volume $V_d$ 
is related to $N_d$ by $V_d=V_d/\bar{n} $ in terms 
of the average droplet density $\bar n$ 
and is proportional to $r_c^3$ for not thin 
droplets.  We also define  the evaporation 
rate near the contact line,   
\be 
W_{\rm con}^e = 2\pi \int_{r_{\rm c}-r_w}^{r_{\rm c}} 
dr {r}J/{\sin\theta}  .
\en 
Here  we set $r_w =6\ell$, which is twice longer than 
the interface width  and remains shorter than 
$r_c$ before the droplet disappearance. In fig.~5, we show 
$W_{\rm tot}^e(t)$ and $W_{\rm con}^e(t)$ versus 
$t$.  The two curves  nearly coincide  for $t \gs 10^3$. 
This demonstrates that  evaporation  
 occurs only near  the contact line at long times.  Also 
in fig.~3, along the interface far from the substrate $z\gg \xi$,   
we can see ${\bi \nu}\cdot\nabla T=0$ and recognize 
 $J=0$ from eq.~(17).  
In terms of $v_g$ in eq.~(14) we then obtain  
\be 
W_{\rm tot}^e 
\sim 2\pi  \xi r_c n_gv_g.
\en 
If the droplet density is treated as  a 
constant($=n_\ell$), eqs.~(14) and (21) 
yield 
\bea 
\frac{d}{dt}  r_c(t)^2 &\sim& 
 -  W_{\rm tot}^e/r_cn_\ell \nonumber\\
&\sim& 
 - \lambda_\ell \Delta T/T_{\rm cx}n_\ell\Delta s    ,
\ena 
which is consistent with the curve in fig.~4 at long times 
$t\gs 10^3$. 
The agreement with the decay behavior  (15) 
becomes better if we  account for 
 the slow  decrease  of the 
the droplet density (see 
the discussion below eq.~(15)).

As shown in the 
 inset of fig.~5, 
$W_{\rm tot}^e$ is much enhanced 
exhibiting  
 oscillatory behavior 
on the acoustic time scale ($\sim $the droplet radius divided 
by the liquid sound velocity), where evaporation 
takes place all over the interface. 
The sound velocity 
is $5.43$ in liquid and $2.67$ in gas 
in units of $\ell/\tau_0$ (with $\sigma=0.06$) at $T=0.875T_c$, 
 so the acoutic disturbances 
 propagate faster in liquid. 
Before the  first peak of  $W_{\rm tot}^e$,  
a compression  sound pulse 
 emitted from the substrate  adiabatically warms 
the liquid  \cite{Miura,Ferrell}, 
resulting in a rapid rise of  evaporation 
on all the interface. 
Its subsequent sharp drop at $t\sim 7$ then 
occurs with propagation of an expansion sound wave 
emitted from the contact line, where  
evaporation suddenly starts and cools  the surrounding liquid. 
These acoustic waves are 
mixed after the first minimum.

\begin{figure}
\begin{center}
\includegraphics[width=0.75\linewidth]{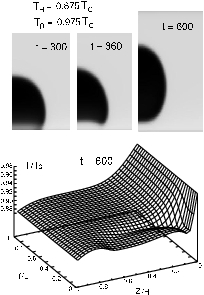}
\end{center}
\caption{\protect
Leidenfrost effect for 
$T_0=0.975T_c$ and 
$T_H=0.875T_c$. Top plates: a gas layer appears 
between the droplet and the substrate (left) and 
the droplet is detached from the substrate (middle and right).  
Bottom plate: Profile of 
$T(r,z,t)/T_c$ at $t=600$ after droplet detachment, 
where $T$ is nearly constant along the interface. 
The gas layer between the droplet and 
the substrate has a large temperature gradient. 
}
\end{figure}

\section{ Leidenfrost effect} 

In fig.~6, we show an extreme 
case of heating 
 the substrate from  $0.875T_c$ 
to $0.975T_c$ at $t=0$. Here, at the pressure in the gas,  
the substrate temperature is much above $T_{\rm cx}(p)$ 
and the thermal diffusion is relatively slow 
near the substrate since $T_0\cong T_c$. 
On a time scale of order $ 100$,  
a gas layer emerges on the substrate 
and it supports a large temperature gradient. 
Here the whole liquid layer adjacent to the heater 
turns simultaneously to the gas. 
The droplet is then detached 
from the substrate (the Leidenfrost effect). 
Without gravity in this case, 
the droplet slowly moves towards 
the cooler boundary and the droplet interior   becomes 
 cooler than the surrounding gas region 
due to evaporation. 
It eventually collides with the 
cooler boundary to form a thickened 
wetting layer (see fig.~7 of our previous 
paper \cite{OnukiV}).      In gravity, a droplet 
may be suspended in gas in a steady state. 
On the other hand, a gas bubble is 
attacted to a warmer boundary \cite{Hegseth,OnukiV}. 
The lower panel of fig.~6 
demonstrates that the  temperature 
is nearly homogeneous all along the 
interface at $T_{\rm cx}(p)$ 
with disappearance of 
 the contact line. Here the temperature 
in the gas layer between the droplet and the 
substrate steeply changes from 
$T_0$ to $T_{\rm cx}(p)$. Thus the layer 
is  strongly absorbing heat from the substrate.

\section{Summary and concluding remarks}

For one-component fluids we 
 have  examined  evaporation of a small droplet 
on a heated smooth substrate  
in the partial wetting condition 
in the axisymmetric geometry. 
In the dynamic van der Waals 
theory \cite{OnukiV}, we have integrated the entropy equation (8) 
together with the continuity and momentum equations 
to avoid parasitic flow around the interface. 
Our system length  
is of the order of 
several ten nanometers 
if the mesh length $\Delta x$ 
 is a few $\rm \AA$.

In our simulation, the temperature 
exhibits a sharp drop  
near the  contact line, 
leading to a large temperature gradient 
and a large heat flux localized near    the contact line. 
As is evident in fig.~5,    
evaporation in one-component fluids 
 occurs only near the contact line at long times.  
Here we should note that we 
have assumed the isothermal 
boundary condition on the substrate.  
For finite  thermal conductivity 
of the wall, however, the substrate temperature  
is lowered at  the contact line and  the temperature 
 drop  to $T_{\rm cx}(p)$   in the fluid  
should become  much more gradual 
   near the contact line \cite{Nikolayev,NikolayevNEW,Stephan}.

Phenomenologically, the evaporation rate of a thin 
liquid droplet 
in air has been assumed to be 
of the form \cite{Nature,Caza,Bonn}, 
\be 
J(r,t) =J_0/\sqrt{r_c(t)^2 -r^2},
\en  
Here $J_0$ is a constant, but 
its expression in terms of the physical 
parameters remains  unknown. 
The total evaporation rate 
 $W_{\rm tot}^e$, which is 
the surface integral of $J(r,t)$,   
is proportional to $r_c$ as in our case in eq.~(20). 
If the temporal decrease of 
the liquid density 
 is neglected, the above $J(r,t)$  
yields $r_c(t) \propto (t_0-t)^{0.5}$ 
 in agreement with the experiments\cite{Nature,Caza,Bonn}. 
This decay law is analogous to 
that in eq.~(21) for one-component fluids, 
although the forms of 
 $J(r,t)$ in the two cases are very different. 
In multi-component 
fluids,     a velocity   field 
induced by the Marangoni  effect 
should  serves to   realize  evaporation on 
all the interfcae\cite{Savino,Larson}. 

\acknowledgments
This work was supported by KAKENHI (Grant-in-Aid for Scientific Research) on Priority Area gSoft Matter Physicsh from the Ministry of Education, Culture, Sports, Science and Technology of Japan.



\begin{thebibliography}{0}



\bibitem{PG}  De Gennes P.G., 
Rev. Mod. Phys. {\bf 57} (1985) 827. 





\bibitem{Nature} Deegan R.D.,  Bakajin O., 
 Dupont T.F., Huber G.,  Nagel S.R. and Witten  T.A., 
Nature {\bf 389} (1997) 827.  

\bibitem{Caza}
  Gu$\acute{\rm e}$na G.,   Poulard C.    
and   Cazabat A.M., Colloid and Interface Science
 {\bf 312} (2007) 164. 
 

\bibitem{Bonn}   Shahidzadeh-Bonn N.,  Rafai S.,  Azouni A.  
and Bonn D., J. Fluid. Mech. {\bf 549} (2006) 307.  


\bibitem{Butt}
Butt H.J., Glovko D.S. and Bonaccurso E., 
J. Phys.  Chem B {\bf 111} (2007) 5277. 




\bibitem{Davis}  Anderson D. M. and  Davis S.H., 
Phys. Fluids, {\bf 7} (1995)  248. 

\bibitem{Koplik}  
  Koplik J.,  Pal S.   and  Banavar J.R.,    
 Phys. Rev. E {\bf 65} (2002) 021504.   



\bibitem{Yeomans}  Briant A. J., Wagner  A.J.  and 
Yeomans J.M.,  Phys. Rev. E {\bf 69} (2004) 031602. 

\bibitem{Savino} Savino R.  and  Fico S., 
Phys. Fluids, {\bf 16} (2004) 3738. 

\bibitem{Larson} Hu H. 
and  Larson R.G., Langmuir {\bf 21} (2005) 3972. 

\bibitem{Nikolayev} 
 Nikolayev V.S., Beysens D.A., 
 Lagier G.L. 
 and Hegseth J.,  
Int. J. of Heat and Mass Transfer {\bf 44} (2001)  3499.  
\bibitem{NikolayevNEW} 
Nikolayev V.S. , preprint ( arXiv:0709.4631).





\bibitem{Hegseth} 
 Hegseth J., Oprisan A., Garrabos Y., 
Nikolayev V.S., Lecoutre-Chabot C.
 and  Beysens D. 
Phys. Rev. E {\bf 72} (2005) 031602. 




\bibitem{OnukiV} 
Onuki A.,   
{Phys. Rev. E} 
 {\bf 75} (2007)  036304.

\bibitem{Kanatani} Onuki A. and  Kanatani K., 
Phys. Rev. E {\bf 72} (2005) 066304. 

\bibitem{Stephan} H$\ddot{\rm o}$hmann C. and Stephan P., 
Exp. Thermal and Fluid Sci., {\bf 26} (2002) 157. 



\bibitem{Landau}   Landau L.D. and  
Lifshitz E.M.,  {\it Fluid Mechanics} (Pergamon, 1959). 

\bibitem{vander} 
Rowlinson J.S.,   J. Stat. Phys. {\bf 20} 
(1979) 197.



\bibitem{para} Lafaurie B., Nardone C., Scardovelli R., 
Zaleski S. and  Zanetti G., 
J. Comput. Phys. {\bf 113} (1994) 134.
  

\bibitem{Jamet1} 
Jamet D,  Torres D  and  Brackbill J.U., 
J. Comput. Phys. {\bf 182} (2002) 262.
\bibitem{Daru} 
 Shin S, Abdel-Khalik S.I.,  Daru V. and 
 Juric D., J. Comput. Phys. {\bf 203} (2005) 493.  





 




\bibitem{Bird} 
Bird R.B.,  Stewart W.E. and 
Lightfoot E.N., {\it Transport Phenomena} (Wiley, New York, 2002), 
p.272.   



\bibitem{Miura} 
Miura Y., Yoshihara S., 
 Ohnishi M.,  Honda K., 
 Matsumoto M.,  Kawai J., 
 Ishikawa M., 
Kobayashi H. 
and   Onuki A., 
 Phys. Rev. E {\bf 74} (2006) 010101(R). 



\bibitem{Ferrell} 
Onuki A., Phys. Rev. E {\bf 76} (2007) 061126.  
 

\end{thebibliography}
\end{document}